\documentclass[final,1p,times]{elsarticle}

\usepackage{graphicx}
\usepackage{subfigure}
\usepackage{amssymb}
\usepackage{amsthm}
\usepackage{lineno}

\journal{Nuclear Physics A}

\begin{document}

\begin{frontmatter}

\title{Measurements of identified particle anisotropic flow in Cu+Au and U+U collisions by PHENIX experiment}

\author{Shengli Huang (for the PHENIX Collaboration)}
\address{Vanderbilt University, Nashville, TN, USA  37240}

\begin{abstract}
In this paper, new measurements of anisotropic flow ($v_{1}$,
$v_{2}$) for identified particles such as pions and protons in
Cu+Au collisions at $\sqrt{s_{NN}}$ = 200 GeV and U+U collisions
at $\sqrt{s_{NN}}$ = 193 GeV are reported. The anisotropic flow is
studied as a function of $p_{T}$ and centrality in these two
collision systems. In ultra-central U+U collisions (0--2\%
centrality), $v_2$ of protons shows a weak $p_{T}$ dependence for
$p_T<$~1.0 GeV/c.  A positive $v_1$ for charged pions is observed
for $p_{T} >$ 1 GeV/c with respect to the first-order event plane,
the angle of which is determined by the Au-going spectators. The
scaling of identified particle $v_2$ with the number of valence
quarks ($n_q$) has been observed in these two collision systems,
with the transverse kinetic energy.
\end{abstract}

\end{frontmatter} 


\section{Introduction}
Measurements of the anisotropic flow with different order harmonic
coefficients ($v_n, n = 1,2,3,4$) have played a pivotal role in
the discovery of quark-gluon plasma (QGP) at RHIC and the
LHC~\cite{whitepaper}. They are also important for the study of
viscous hydrodynamics and the extraction of the shear viscosity
over entropy density ($\eta/s$)~\cite{eta}. Anisotropic flow is
strongly coupled to the medium density, initial geometry shape,
and corresponding event-by-event fluctuations, which are crucial
for further accurate measurements of elliptic flow and
understanding the QGP properties ~\cite{fluctuation}.

The flexibility of RHIC can provide different kinds of heavy ion
collisions in addition to Au+Au and Cu+Cu collisions, which
include the colliding of two deformed nuclei, such as U+U at 193
GeV, and species-asymmetric collisions, such as Cu+Au at 200 GeV.
In central U+U collisions, the initial geometry and eccentricity
will be quite different for tip-tip and body-body
collisions~\cite{UU}. Additionally, Cu+Au collisions can provide
an asymmetric geometry and density both in the transverse plane
and longitudinally~\cite{CUAU}. Measurements of these systems will
open new windows to investigate the influence of different initial
geometry and densities.

In this paper, I will present new measurements of identified
particle $v_{2}$ in U+U collisions at 193~GeV and identified
particle $v_{1}, v_{2}$ in Cu+Au collisions at 200~GeV. The number
of valence quark ($n_{q}$) scaling is also tested for $v_{2}$ in
the Cu+Au and U+U collisions.

\section{Analysis Methods}
In 2012, PHENIX recorded 3.0~B and 4.6~B minimum-bias events in
U+U at 193~GeV and Cu+Au collisions at 200~GeV, respectively. For
the results shown in this paper, 70~M and 600~M events are used
for U+U collisions and Cu+Au collisions respectively.

The $v_n$ measurement is performed by correlating the particle
azimuthal angle $\varphi$ with the corresponding harmonic
event-plane angle $\Psi_{n}$, and correcting the observed signal
for the event-plane resolution as follows:

\begin{equation}
v_n = \frac{\left\langle\cos(n(\varphi -\Psi_n))\right\rangle}
{Res(\Psi_n)} \label{v2def}
\end{equation}
Here the brackets $\langle \rangle$ indicate an average over all
particles in all events for a certain centrality bin and
$Res(\Psi_{n})$ indicates the correction factor for the
event-plane resolution.

The SMDs (Shower Maximum Detectors)~\cite{SMD}, with a rapidity
coverage of $|\eta|\ge$ 6.5, are used to measure the angle of
$\Psi_{1}$. The BBCs~\cite{BBC} (3.1$ \le|\eta|\le$ 3.9) and
MPCs~\cite{MPC} (3.1 $\le|\eta|\le$ 3.9) are used to measure the
angle of $\Psi_{2}$ in Cu+Au and U+U collisions respectively. Two
TOF (Time-of-Flight) detectors~\cite{BBC,TOF} which sit in the
west and east arms of PHENIX were used to identify the particles.
The pions, kaons and protons can be separated up to $p_{T}\sim$ 3
GeV/c by TOF with a timing resolution of 89 ps (TOFw) and 120 ps
(TOFe), respectively.

\section{Results and discussions}

\subsection{$v_{2}$ of pions and protons in U+U collisions at 193 GeV}

\begin{figure}[htbp]
\begin{minipage}[t]{14pc}
\includegraphics[scale=0.31]{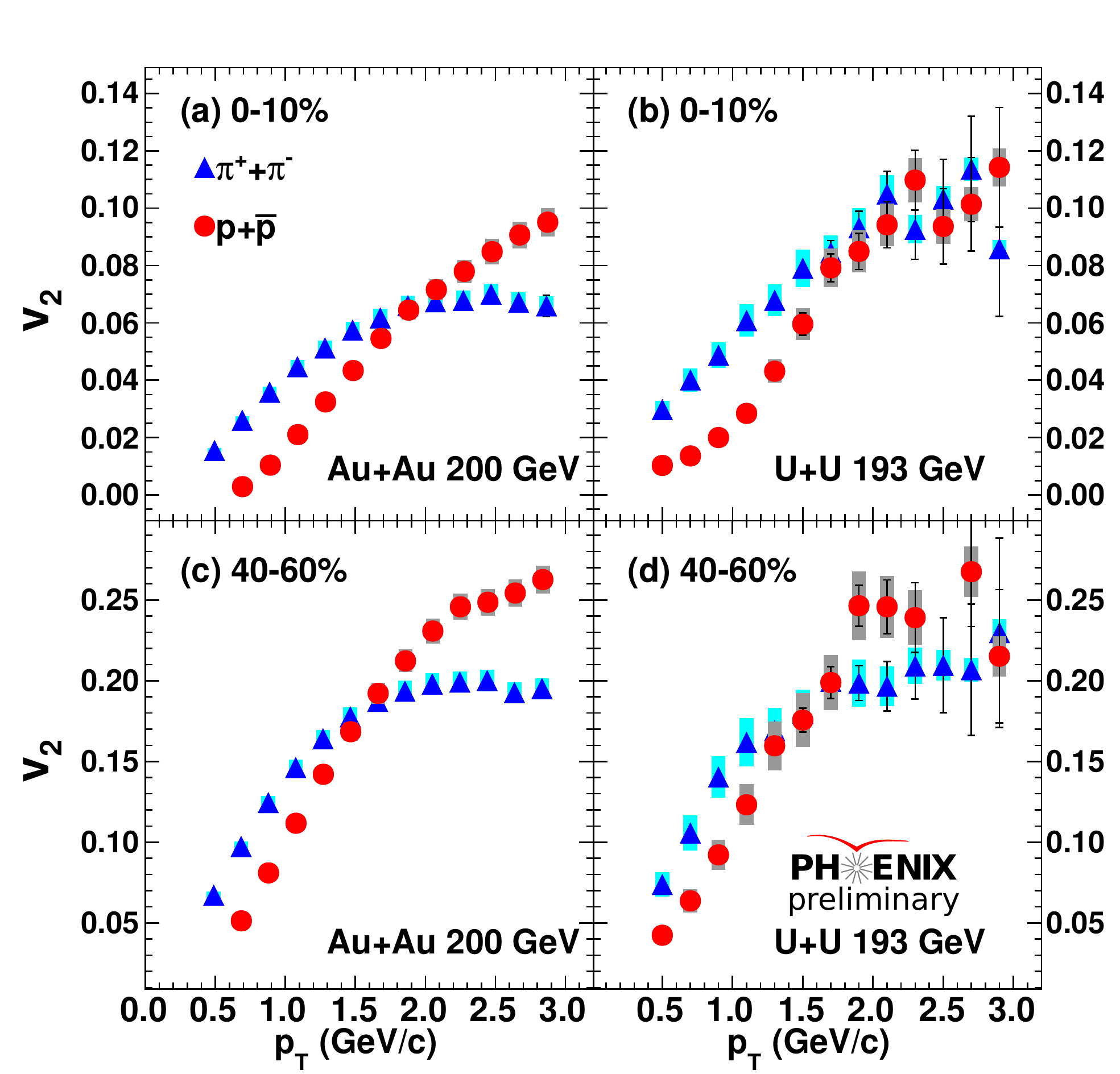}
\caption{$v_{2}$ of pions and protons as a function of $p_{T}$ for
the centrality bins of 0--10\% and 40--60\% in Au+Au at 200 GeV
and U+U collisions at 193 GeV.} \label{fig:widebin_uu}
\end{minipage}\hspace{2pc}
\begin{minipage}[t]{14pc}
\includegraphics[scale=0.31]{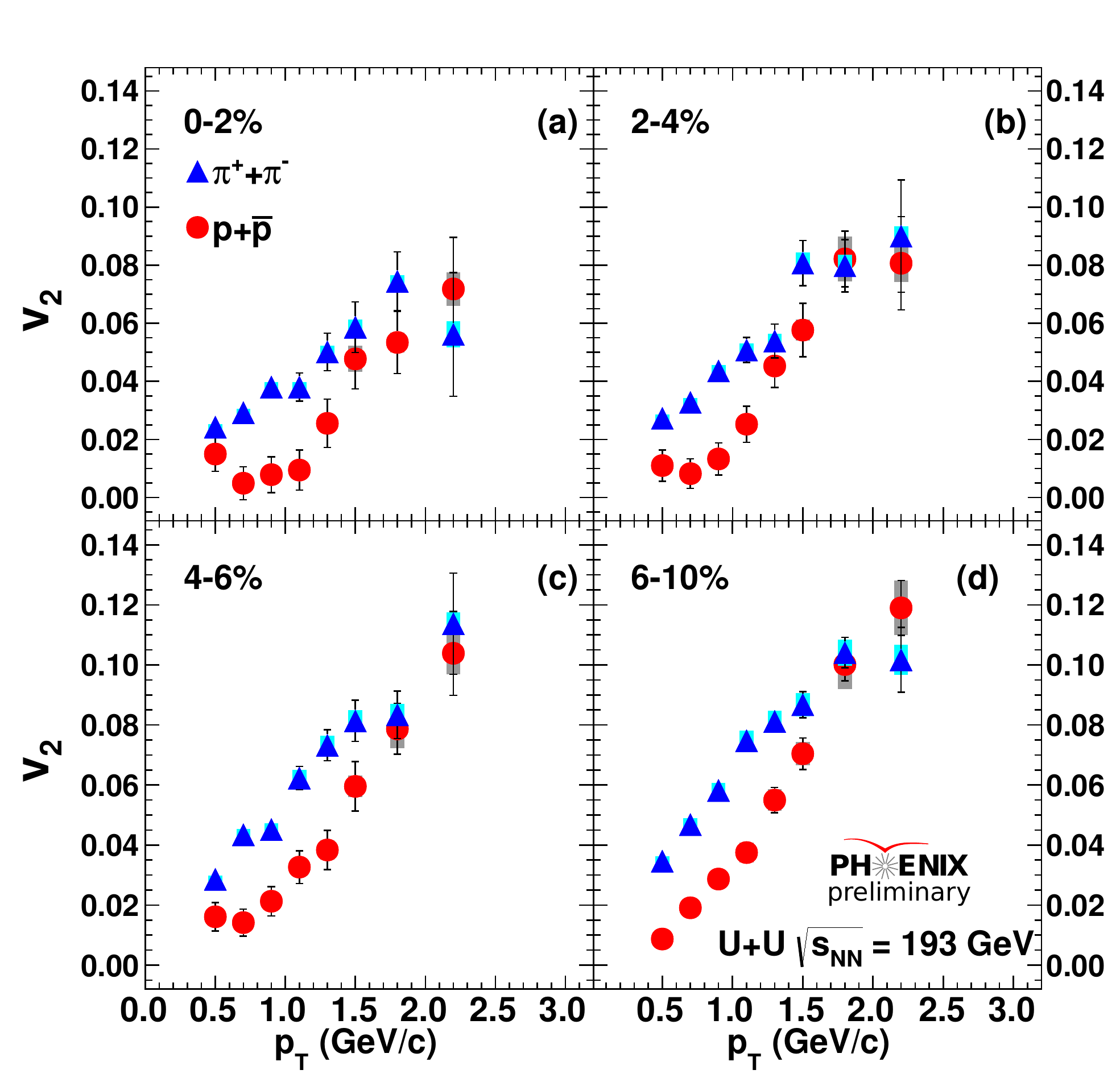}
\caption{$v_{2}$ of pions and protons as a function of $p_{T}$ for
the finer centrality bins of 0--2\%, 2--4\%, 4--6\% and 6--10\% in
U+U collisions at 193~GeV.} \label{fig:finerbin_uu}
\end{minipage}
\end{figure}

The results for $v_2$ of pions and protons in U+U collisions at
193 GeV are presented in Fig.\ref{fig:widebin_uu}; As
Fig.\ref{fig:widebin_uu} shows, the protons show a weaker $p_{T}$
dependence in 0--10\% centrality bin below $p_{T}<$ 1~GeV/c,
compared with that of Au+Au collisions~\cite{TOF}. However, the
$p_{T}$ dependence of $v_{2}$ of pions, and for both particles in
non-central collisions are quite similar between Au+Au and U+U
collisions.

To further investigate this behavior, the $v_{2}$ of pions and
protons are studied in finer centrality bins such as 0--2\%,
2--4\%, 4--6\% and 6--10\% in U+U collision, with centrality being
defined by the total charge signal from the BBCs. As
Fig.\ref{fig:finerbin_uu} shows, the weak $p_{T}$ dependence for
$v_{2}$ of protons below $p_{T}<$ 1~GeV/c grows more pronounced in
the more central U+U collisions such as 0-2\%. In the future, we
will try to separate the collision geometry into tip-tip and
body-body collisions, so that we can further understand what cause
this behavior. Possible explanations are stronger radial flow due
to a high medium density or the complicated initial geometry in
central U+U collisions.

\subsection{$v_{1}$ of charged pions in Cu+Au collisions at 200~GeV}
\begin{figure}[htbp]
\begin{center}
\includegraphics[scale=0.31]{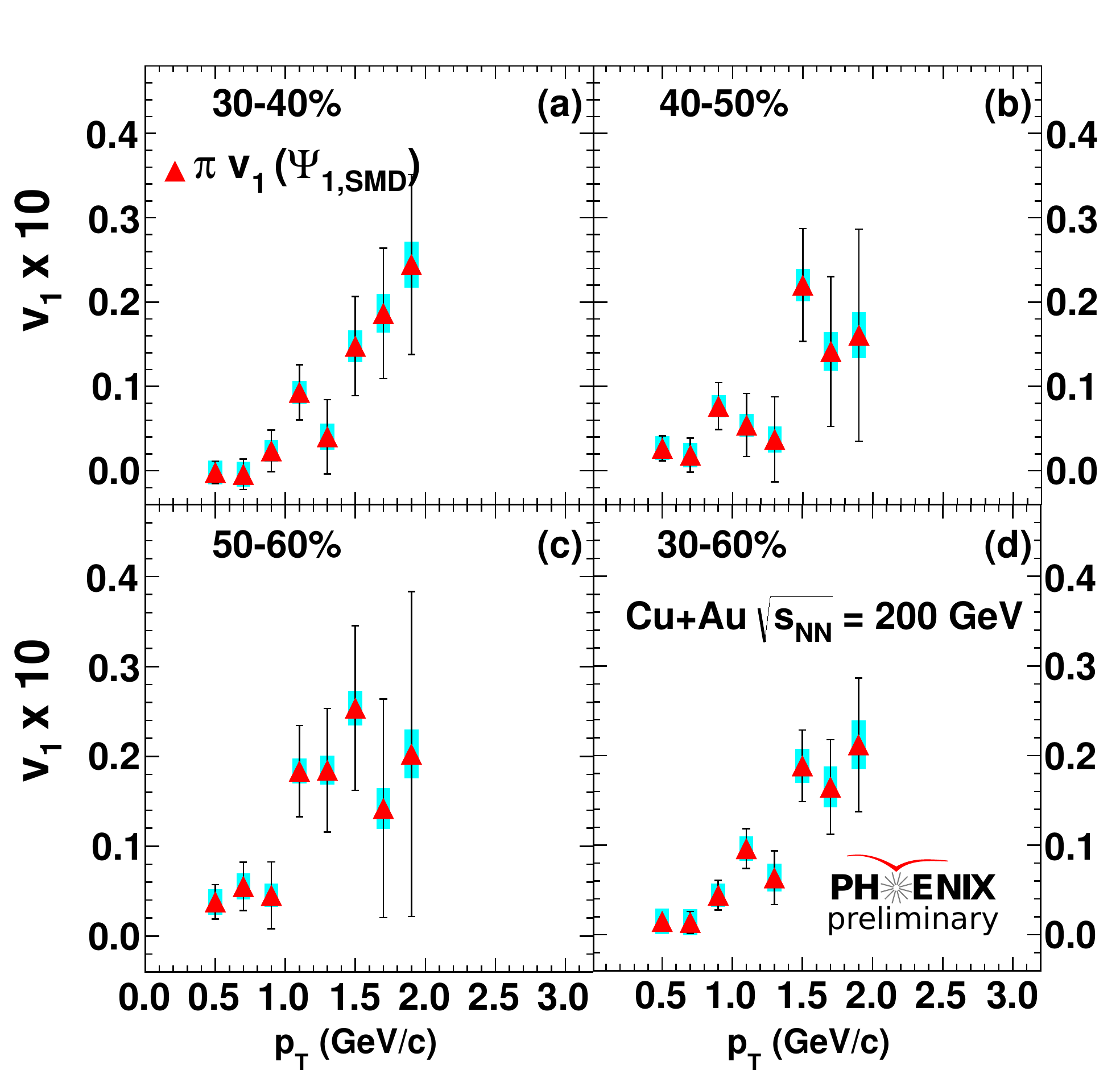}
\end{center}
\caption{$v_{1}$ of pions as a function of $p_{T}$ for centrality
bins of 30--40\%, 40--50\%, 50--60\% and 30--60\% in Cu+Au
collisions at 200~GeV. The angle of $\Psi_{1}$ is determined by
the Au-going spectators which are measured by SMD.}
\label{fig:pionv1_cuau}
\end{figure}

In asymmetric collisions such as Cu+Au, the asymmetric initial
density profile may generate an asymmetric pressure gradient with
respect to the reaction plane, which will lead to an asymmetric
momentum space distribution with respect to the reaction plane.
The partons generated by hard scattering will lose different
amounts of energy, due to the fact that they are emitted from
different medium surfaces, either on the Au-going side or the
Cu-going side, which will also generate an asymmetric momentum
space distribution at high $p_{T}$. Therefore, the measurements of
first order harmonic coefficients ($v_{1}$) in a broad $p_{T}$
region in Cu+Au will supply more information about this asymmetric
medium. The $v_{1}$ of charged pions as a function of $p_{T}$ and
centrality is shown in Fig.\ref{fig:pionv1_cuau}. A positive
$v_{1}$ is observed for $p_{T}\ge$~1~GeV/c. The angle of
$\Psi_{1}$ is determined by the Au-going spectators, so a positive
$v_{1}$ measured by Equ.~\ref{v2def} indicates that there are more
particles emitted from the Au-going side at $p_{T}\ge$~1~GeV/c.

\subsection{$n_{q}$ scaling in Cu+Au and U+U collisions}
\begin{figure}[htbp]
\begin{center}
\includegraphics[scale=0.31]{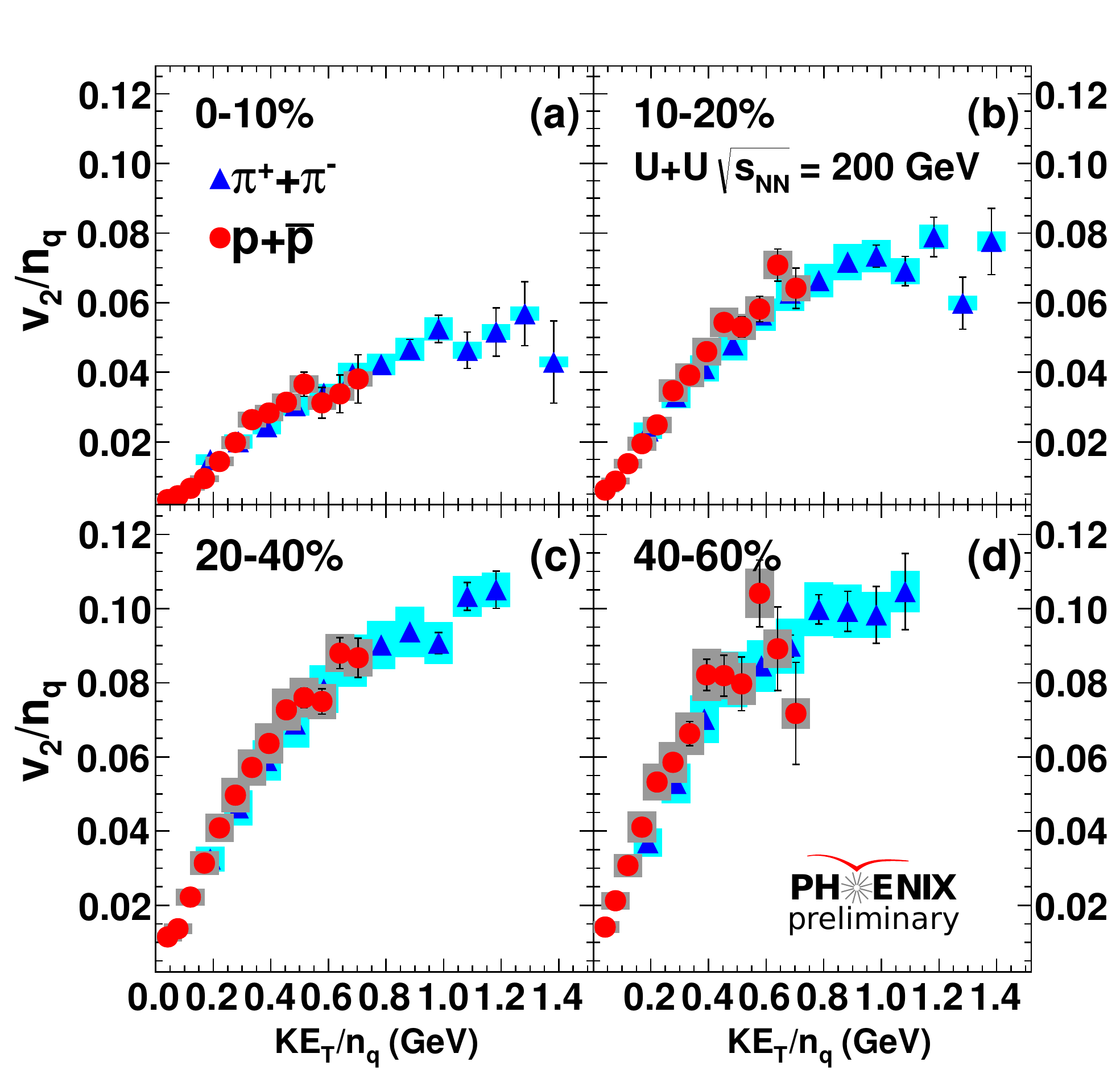}
\includegraphics[scale=0.31]{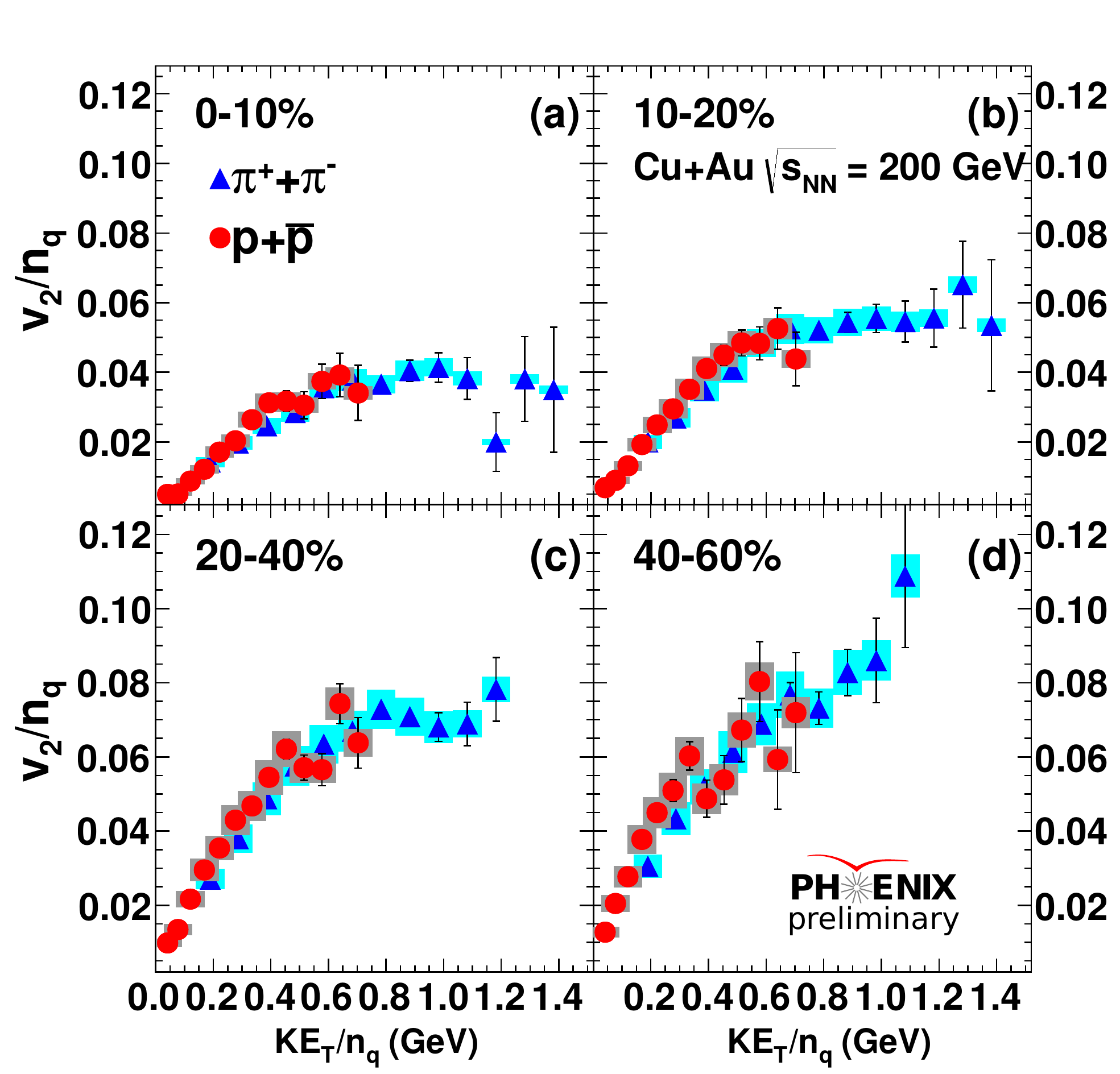}
\end{center}
\caption{$v_{2}/n_{q}$ of pions and protons as a function of
$KE_{T}/n_{q}$ for centrality bins 0--10\%, 10--20\%, 20--40\% and
40--60\% in U+U collisions at 193 GeV (left panel) and Cu+Au
collisions at 200 GeV (right panel). The error bars (shaded boxes)
represent the statistical (systematic) uncertainties.}
\label{fig:nq-scaling}
\end{figure}
The observation of $n_q$ scaling has been the basis of the claim
that a partonic matter with quark-like degrees of freedom and
significant collectivity has been generated in heavy ion
collisions~\cite{nqscaling}. In U+U collisions, the $n_q$ scaling
may be affected by the stronger radial flow, higher medium
density, and complicated initial geometry. In Cu+Au collisions, it
is also interesting to determine whether there are cold nuclear
matter effects, which may give extra contributions to the particle
production. We measured the $n_{q}$ scaling in U+U and Cu+Au
collisions in different centrality bins, which are shown in
Fig.\ref{fig:nq-scaling}. Due to statistical limitations, the
measurement is limited to the region of $KE_{T}/n_{q}\le$ 0.8 GeV,
where the $n_{q}$ scaling is still observed to hold.

\section{Summary}
The $v_{1}$ and $v_{2}$ of pions and protons are measured in U+U
collisions at 193~GeV and Cu+Au collisions at 200~GeV. The $v_{2}$
of proton shows a weak $p_{T}$ dependence for $p_T<$~1~GeV/c in
0--2\% ultra-central U+U collisions. Additionally, a positive
$v_{1}$ of charged pions is observed for $p_{T}\ge$~1~GeV/c. These
new measurements will give us further information to understand
the initial geometry, medium density, and their corresponding
fluctuations in heavy ion collisions.


\end{document}